# Spatio-spectral light-by-light moulding in multimode fibre

**Yago Arosa[1,3,†], Tigran Mansuryan[1,†], Arnaud Poisson[1], Wasyhun Asefa Gemechu[2], Katarzyna Krupa[4], Mario Ferraro[2], Fabio Mangini[2], Benjamin Wetzel[1], Stefan Wabnitz[2], Alessandro Tonello[1,‡], Vincent Couderc[1,‡,*]**

[1]*Université de Limoges, XLIM, UMR CNRS 7252, 123 Avenue A. Thomas, 87060 Limoges, France*
[2]*DIET, Sapienza University of Rome Via Eudossiana 18, 00184 Rome, Italy*
[3]*iMATUS, University of Santiago de Compostela, Praza do Obradoiro S/N, Santiago de Compostela, Coruña 15782, Spain.*
[4]*Institute of Physical Chemistry, Polish Academy of Sciences, ul. Kasprzaka 44/52, 01-224 Warsaw, Poland*

e-mail* Vincent.couderc@xlim.fr
[†] These authors contributed equally to this work and share first authorship
[‡] These authors contributed equally to this work and share senior authorship

**Controlling complex light waves to achieve desired behaviours or characteristics on demand presents a significant challenge. This task becomes even more complicated when manipulating speckled light beams owing to their inherently fuzzy intensity and phase structures. Here, we demonstrate that a weak speckled second-harmonic signal in a multimode graded-index fibre can be manipulated via its conservative interaction with a high-power co-propagating fundamental pump wave. Specifically, the spatial quality of the signal can be either enhanced or degraded by varying the pump's power or its modal power distribution. The underlying physical mechanism is the optically induced mode conversion, whose phase-matching can be controlled by the mode power distribution of the pump beam. This phenomenon enables new possibilities for manipulating complex light via material nonlinearities in multimode guiding structures. A striking example of this novel light-by-light control is the experimentally observed enhancement or partial suppression of the visible Raman Stokes cascade regulated by the second harmonic beam, while modulated by the mode power distribution of the fundamental beam.**

The manipulation of light-by-light through cross-phase modulation (XPM) via the Kerr effect was extensively studied in the 1980s and 1990s [1-7]. The Kerr-induced cross-action of an intense pump light beam can alter a weak copropagating probe beam's temporal, spectral and spatial profiles. This interaction occurs conservatively without any power exchange between the pump and the probe. For instance, the additional phase shift that is induced by the strong wave may enable phase-matching between a fundamental frequency (FF) beam and its second harmonic (SH), allowing the control of nonlinear frequency conversion processes [8]. A lesser-known method for conservative control between light beams of different frequencies is the optically induced mode coupling (OIMC). This parametric interaction permits frequency translation without exponential gain, thereby avoiding spontaneous noise emission. As a result, the frequency translation associated with OIMC preserves the quantum state of a signal [9-10]. OIMC can involve two pumps at different wavelengths: maximum conversion efficiency from a signal to an idler is obtained at phase matching, under various configurations of polarization states, spatial modes, and wavelengths for pumps, signals, and idlers [9,11-12].

Nonlinear effects in multimode optical fibres provide a leap in our capacity for all-optical light manipulation: recent experiments have revealed that controlling the input wavefront permits to shape the spatial, temporal, and spectral properties of highly multimode beams [13-14]. In this framework, the effect of Kerr beam self-cleaning (BSC) [15-26], which allows the transformation of a speckle into a quasi-single-mode beam at the output of a graded-index (GRIN) highly multimode optical fibre (MMF), is particularly intriguing. BSC was initially interpreted as a nonlinear interaction between linear beam self-imaging and Kerr nonlinearity. The resulting nonlinear refractive index long-period grating [15-16] facilitates the phase-matching of frequency degenerate inter-modal four-wave mixing (IM-FWM), thus enhancing the nonlinear power redistribution [16-18]. From the application point of view, the most important aspect of BSC is the enhancement of the beam spatial quality, that is, the quenching of the beam's $M^2$ factor. Although it is a purely conservative process, BSC has also been observed in lossy and gain fibres [24-25] and in active tapers [27-28]. This has opened a new avenue for developing powerful fibre lasers operating in Q-switched or mode-locked configurations [25-26, 29-32]. Experiments have shown that BSC maintains the spatial coherence of the input laser [33], which enables new applications, such as imaging through MMFs [34]. Nevertheless, explaining the underlying mechanism for BSC has led to an ongoing controversial debate. Currently, the most widely accepted theory of BSC is based on thermodynamic considerations [35-37]. Within such a theoretical framework, the concept of BSC has been extended to the nonlinear dynamics of two co-propagating, orthogonally polarized beams [38-41]. However, to our knowledge, experimental demonstrations of BSC involving multiple beams have been limited to a single carrier wavelength.

In this paper, we elevate the field of light manipulation in MMFs to a new dimension, involving the nonlinear interaction among multiple laser beams of different carrier wavelengths. We experimentally demonstrate that the evolution of a weak speckled beam in the visible range can be controlled using a combination of XPM and nondegenerate OIMC with an intense pump beam in the near infrared, which undergoes its own spatial reshaping through the Kerr BSC process. In our experiments, the visible green beam is the SH of a Yb laser, which is used as the FF infrared control beam. Both beams are simultaneously injected into a GRIN MMF. By varying the power and modal distribution of the FF beam, the M² quality factor of the visible SH beam at the MMF output can be either decreased (beam cross-cleaning, BXC) or increased (beam cross-spoiling, BXS). The associated modification of the mode power distribution of the SH beam results from phase-matched IM-FWM-induced energy exchanges between its transverse modes, driven by the intense FF beam. This mechanism acts as an active diffuser, which controls the direction of power flow towards high- or low-order modes via the FF pump power. In our first series of experiments, the effect was limited by dispersive temporal walk-off because of the relatively short pulses involved. In a second set of experiments, using long pulses, we show that modal cross-interactions may have a dramatic impact when both beams propagate in the highly nonlinear regime, leading to a Raman cascade of Stokes lines. In this case, our experiments reveal that it is possible to reshape the number and modal content of the Stokes lines generated by a visible light beam by controlling the spatial profile and power of the near-infrared beam.

We validated our observations by nonlinear simulations, including spatial and temporal dimensions. Since the process depends on instantaneous peak power, the transition between BXC and BXS states occurs in an ultrafast regime and can be fully described by the rising edge of a Gaussian pulse. Another simple model that shows the basic physical mechanisms behind the energy exchanges between each beam of the same color is developed in the supplementary information SI1. Nonlinear shaping of the spatial profile of a laser beam using another laser beam paves the way for applications such as beam switching or manipulation in ultrafast regimes, e.g., for deep biological imaging.

**Experimental results**. In our experiments (Figure 1(a)), a Gaussian beam composed of 100 ps chirped pulses at 1030 nm is first split into its two orthogonal polarization components with adjustable powers (Figure 1(a)). The vertical component of the fundamental frequency (FF) pump beam is converted to its second harmonic (SH) by a type I BBO crystal positioned for strong phase mismatch conditions; the unconverted portion of the FF is filtered out. Next, the SH is temporally and spatially superimposed with the horizontal component of the FF beam by a beam splitter and a half-waveplate, before being coupled into the 50/125 µm GRIN fibre. Under these conditions, both beams retain horizontal linear polarization, and their coupling efficiencies into the fibre can be adjusted. The green beam at 515 nm is weak enough, so that its own propagation in the fibre is purely linear. On the other hand, the pump beam may undergo spatial BSC thanks to its high peak power (see supplementary information SI2). As a result, the transverse intensity profile of the pump beam at the fibre output evolves into a clean bell shape, which is observable in both near and far fields. By adjusting the delay line between the visible SH weak beam and the strong FF pump, so that the corresponding pulses no longer overlap in time, we verified that the weak SH beam remained unchanged, even in the presence of the FF beam. Conversely, when pulse synchronization among FF and SH pulses is well-tuned, we observed significant changes in the spatial distribution of the SH at 515 nm. These could be controlled by varying the power of the FF pump beam at 1030 nm, or its spatial coupling conditions. To further investigate this process, we analysed the near (Figures 1(c) and 1(d)) and far fields of the weak SH beam and performed numerical processing of both fields. This allowed the calculation of the beams' spatial phase and M² coefficients, helping to identify their modal energy distribution (Figures 1(b,e,f), respectively).

For the first setting of the input coupling conditions of the FF beam into the fibre (i.e., straight on axis), we observed that increasing the FF power led to mode power redistribution for the SH, whereby lower-order modes (LOMs) of the SH are favoured with respect to higher-order modes (HOMs). This spatial beam reshaping, or BXC, is obtained by the OIMC process, which does not lead to energy exchange between the FF and the SH beams. Our analysis of near and far-field images at the fibre output confirms the beam cross-shaping. The M² coefficient of the SH beam, calculated by standard definitions [42], followed the same trend, decreasing along both the X and Y directions as the FF power grew larger (Figure 1(b), right). The maximum FF peak power in our experiment was approximately 40 kW, limited by input face fibre degradation. This power value remains far below levels typically associated with self-focusing effects.

We could reverse the beam interaction dynamics by slightly adjusting the FF beam's input conditions (i.e., its focus point). Increasing the FF peak power led to a mode power redistribution for the SH, favouring HOMs. This BXS of the SH beam was accompanied by a corresponding increase of its M² coefficient (Figure 1(b), left). Near and far-field analyses confirmed this opposite trend. As we shall see in the subsequent theory section, the redistribution of energy between LOMs and HOMs of the SH beam is driven by the mode power distribution of the pump beam. That is, the FF beam modifies via OIMC the mode content of the weak SH beam. This may change the direction of energy flow among the modes of the SH beam, as testified by the corresponding decrease or increase of its quality factor. The occurrence of BXC or BXS in the SH beam is governed by nonlinear phase shifts, which either preserve or break phase-matching in the OIMC process, respectively.

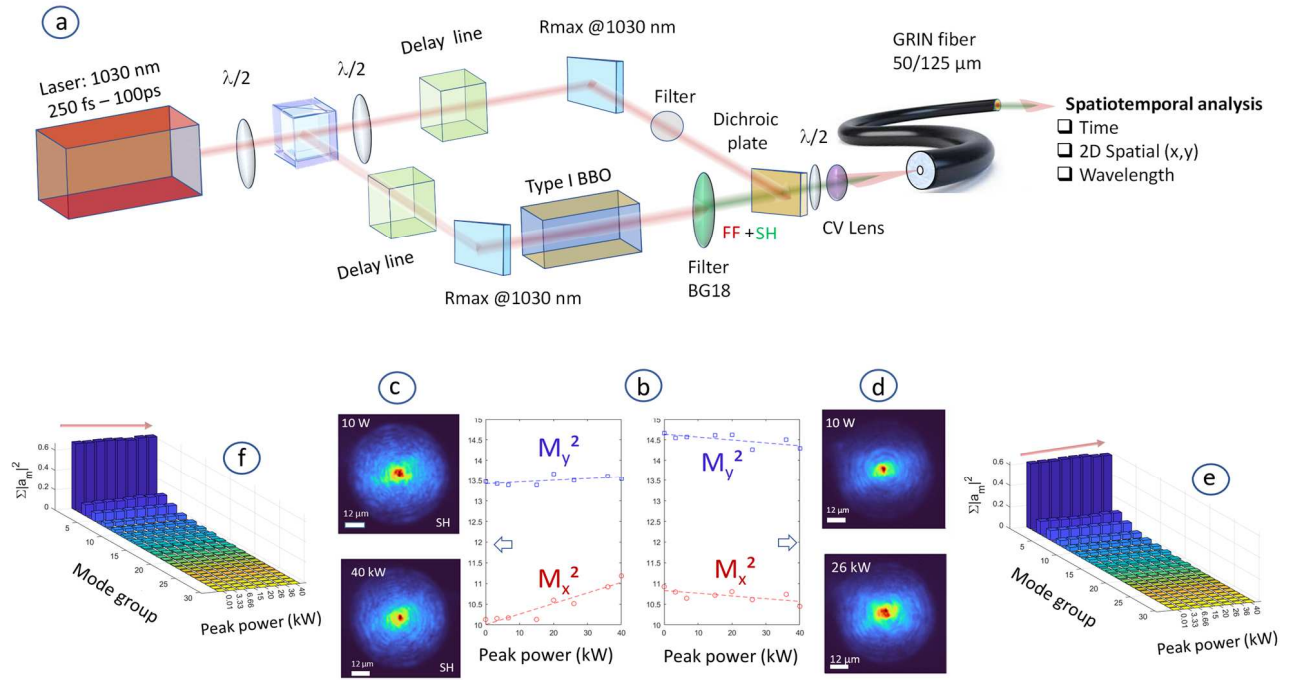

*Figure 1: (a) Experimental setup of spatial cross-manipulation of light by using an FF pump beam and its weak SH beam in a GRIN MMF. (b) M² evolution of the SH beam vs. input infrared pump power for two different cases leading to either (left panel) BXS or (right panel) BXC; (c)(d) examples of output near-field for the (c) BXS or (d) BXC process; (e) energy per mode group for BXC; (f) energy per mode group for BXS.*

In a second series of experiments (Figure 2), we excited additional modes in the FF to achieve a better spatial overlap with the SH. In this scenario, BSC efficiency for the FF beam was reduced, a bell-shaped beam is no longer formed at the fibre output (Figure 2 (a)). Nevertheless, by increasing the FF power we observed a gradual increase in the fundamental mode population of the SH beam (see Figure 2(b) and 2(c)). Maximum SH BXC was observed for input FF peak powers close to 23 kW, leading to a corresponding reduction of the M² coefficient for the SH beam. However, the process reversed at higher input FF powers (up to 100 kW), even without altering the input coupling conditions. Namely, the proportion of energy in the fundamental and other LOMs of the SH beam now decreases as the FF power grows larger (Figure 2(b)). In this regime, HOMs of the SH beam get progressively more populated, which can be interpreted as BXS, leading to an increased SH beam divergence at the fibre output. It is important to note that, to increase the peak power up to 100 kW, it was necessary to reduce the pulse duration from 100 ps down to 10 ps in order to avoid fibre damaging.

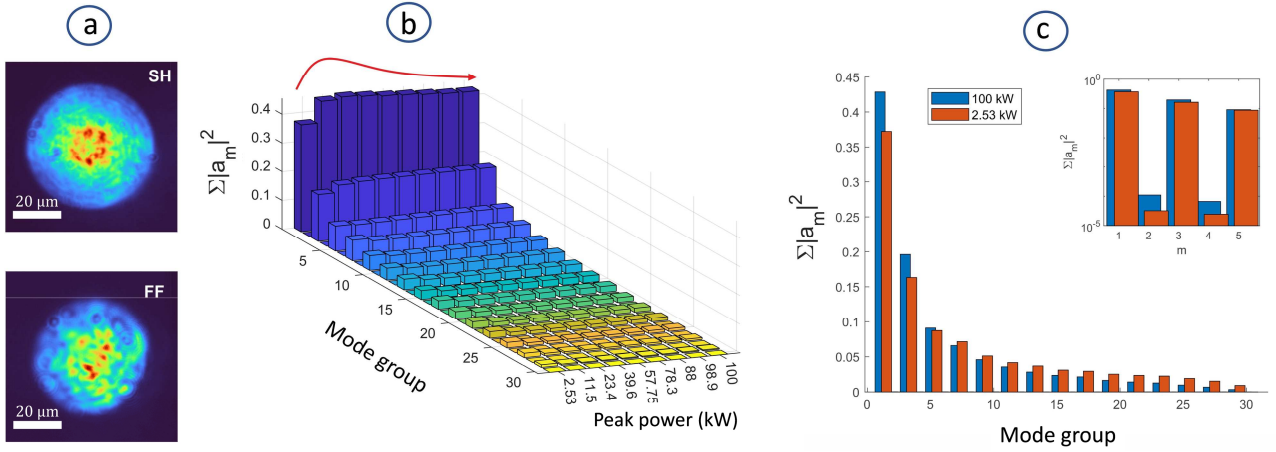

*Figure 2: (a) output near-field transverse intensity profile of FF pump and SH beams; (b) energy distribution in different mode groups of the SH weak beam vs. input peak power of the infrared pump FF, while keeping their input fibre coupling unchanged; (c) SH beam mode energy distribution at fibre output, with either 2.53 kW (orange bars) or 100 kW (blue bars) peak power pulses in FF beam; inset shows that odd mode groups remain weakly populated at all FF powers because of the coupling conditions of the FF beam straight on the fibre axis.*

As we have seen, the BXC efficiency can be limited either by the input FF peak power or by the temporal walk-off between FF and SH pulses. To further investigate the impact of BXC on multimode propagation, we conducted a third series of experiments where both individual SH and FF beams propagate in the fully nonlinear regime. We used a longer GRIN fibre (50 m) with a smaller core diameter (27 µm); to minimize walk-off between the FF and the SH, we used another laser source that emits longer pulses, with a duration of 650 ps. By increasing the power of both SH and FF beams, while maintaining a peak power imbalance of 12:1 in favour of the infrared pump, we observed stimulated Raman scattering (SRS) cascades at both wavelengths, involving the generation of multiple Stokes lines (see Figure 3). The FF beam broadened spectrally beyond 1.3 µm, owing to Raman-shifted temporal soliton dynamics. On the other hand, SH propagation only produced Raman Stokes lines, because of the normal dispersion regime at the SH. Additionally, for both beams nonlinear propagation leads to both spatial BSC and Raman beam clean-up, which concentrate the beam energy into their LOMs (not shown here).

Nondegenerate OIMC between FF and SH beams established a direct link between the visible and fundamental Raman cascades. This interaction allowed the visible Raman cascade to be either enhanced by the BXC (Figures 3(a) and 3(b)) or suppressed (Figures 3(c) and 3(d)) by the BXS effects, induced by the FF pump beam before its spectral broadening. Mutual beam interaction dynamics depends on the intensity, polarization, and input coupling conditions of the infrared FF beam. The observed nonlinear beam dynamics in the presence of the Kerr effect and SRS resembles the behaviour observed in the pure Kerr case. Now, BXC redistributes the energy among the transverse modes of the SH beam, which controls its Raman gain and the associated Stokes cascade. This allowed us to determine the mode content of the beam by separating its spectral components by means of a grating, as depicted in Figure 3(e). Interestingly, we observed that the oscillating transverse mode of the Raman cascade could change (Figures 3(f) and 3(g)) and that the number of Stokes lines could be decreased with a change in the transverse mode (Figures 3(h) and 3(i)). This light-by-light switching was observed across varying numbers of Stokes lines, demonstrating that spatial reshaping of the SH beam preceded the generation of its own Stokes lines. We also note that the modal cascade with the appearance of the $LP_{02}$-like mode remains stable over all three Stokes lines, apparently blocking or stabilizing the Raman beam clean-up process (Figure 3(f) and 3(g)).

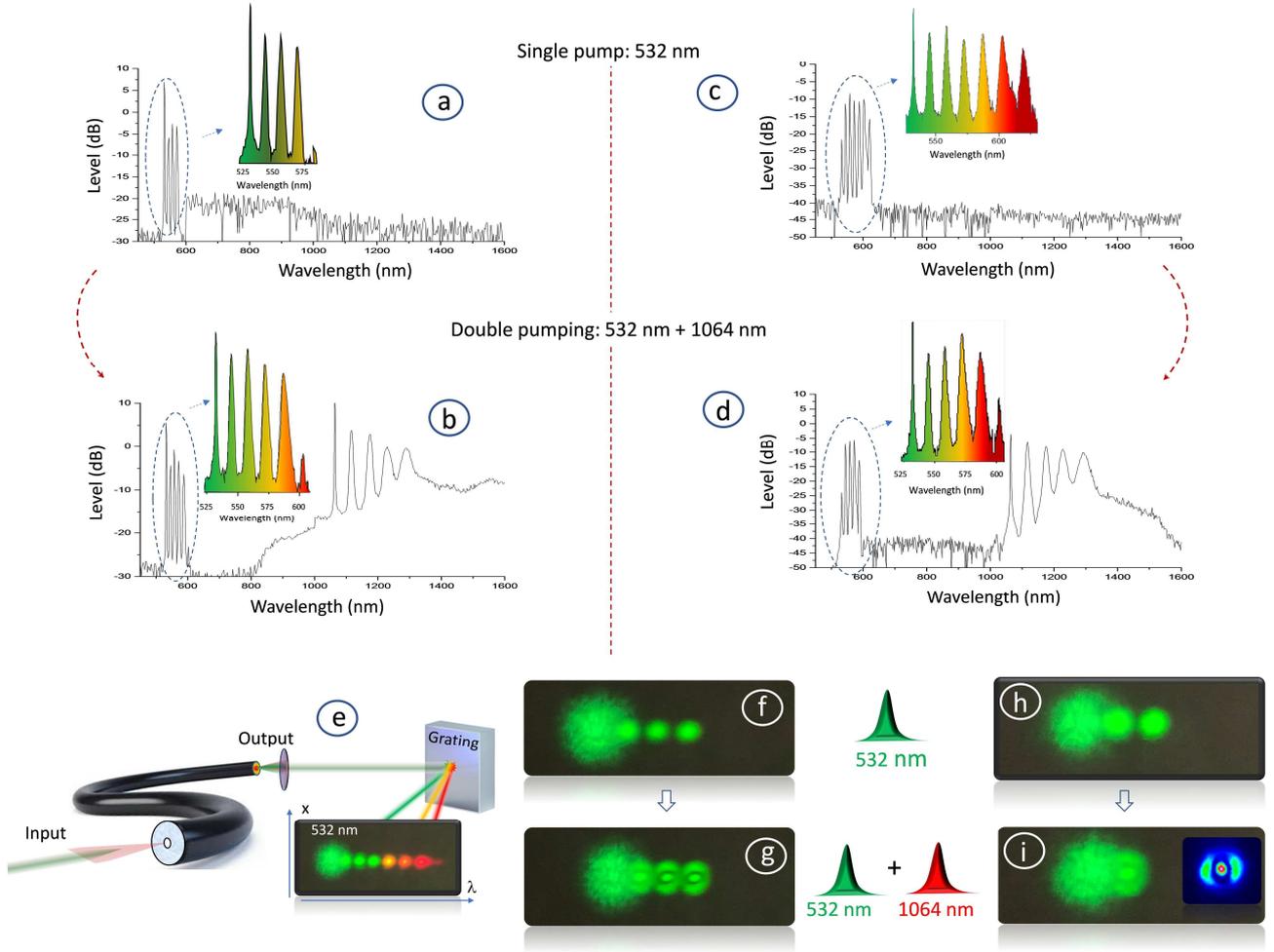

*Figure 3: Experimental results on the cross-action among the FF and the SH beam, when both propagate in a highly nonlinear regime. Fibre length 50 m, GRIN fibre with 27 μm diameter, imbalance of peak power between infrared FF (3kW) and SH beams (250 W), ratio 12:1. SH Raman cascade without (a) or with (b) FF beam, demonstrating BXC-assisted enhancement of Stokes wave generation; SH Raman cascade without (c) or with (d) FF beam, demonstrating BXS-assisted suppression of Stokes wave generation; (e) experimental setup used to simultaneously observe the output beam in wavelength and spatial dimensions. Example of modal switching of the SH: beam profile without and with FF beam, panels (f) and (g), respectively. Example of modal switching and BXS of the SH beam without and with FF beam (h), (i), respectively.*

**Numerical results.** We conducted numerical simulations to analyse the spatial reshaping of the weak SH beam under BXC or BXS, influenced by the FF beam. We used two coupled equations, one for the FF and one for the SH beam, which interact via asymmetric OIMC and XPM, owing to their different wavelengths, powers, and temporal walk-off between their pulses (see Methods). We also included the linear random mode coupling terms. In the linear propagation regime, no evolution of the beams was observed; all excited modes retain their energy, and the $M^2$ coefficient remains constant for both beams (see Supplementary Information SI3a). Simulations show the presence of a spatial BSC process (Figure 4) in the pump beam whenever its power grows above a certain threshold value. As a result, the weak SH beam undergoes BXC, with spatial remodelling leading to power transfer into its fundamental mode at the expense of HOMs (Figure 4(f)). This BXC occurs without any energy exchange between the two beams: no second-order nonlinearity is included in the model. Moreover, when the input peak power of the FF beam is further increased, we observed that BXC starts to saturate, until it gradually reverses (Figure 5). In this regime, the LOMs of the SH beam are now depleted in favour of its HOMs (Figure 5(f)). This inversion of behaviour occurs without any modification of the spatial input coupling conditions of the FF or the SH beams, at variance with the experimental results that we illustrated in Figure 2. Here, the only parameter that varied in our simulations was the FF input peak power. It is important to note that the brightness improvement of the FF pump beam associated with BSC remains limited when compared with the brightness improvement of the weak SH beam under BXC.

We considered input Gaussian beams for both the FF and the SH, centred on the fibre axis in simulations. As a result, only even modes are excited (Figure 4(c)(h) and Figure 5(c)(h)). In the occurrence of both BXC and

BXS of the weak SH beam, we observed that odd modes remain largely unpopulated; this phenomenon has also been highlighted in the simpler case of BSC, as noted in [43]. The authors identified that mode groups containing radially symmetric modes are more populated than odd modes, a natural consequence of injecting a radially symmetric coherent beam into the MMF. They also noted that nonlinear modal interactions have varying efficiencies, due to the significant differences in nonlinear overlap coefficients between modes, which significantly affect the rate of convergence toward a steady-state mode power distribution.

Temporal walk-off between the weak SH and the infrared FF beam plays a significant role in the mechanism of BXC or BXS. To evaluate this, we ran numerical simulations without temporal walk-off (see Supplementary Information SI3b). The results revealed a significant increase in modulation from the infrared to the green beam, along with a more pronounced change in the M² coefficient. Experimentally, temporal walk-off can be minimized by using two closely spaced wavelengths, or a longer pulse relative to the propagation time in the fibre, as realized in our third experiment in strong nonlinear propagation regime involving a remodelling of the Raman cascade.

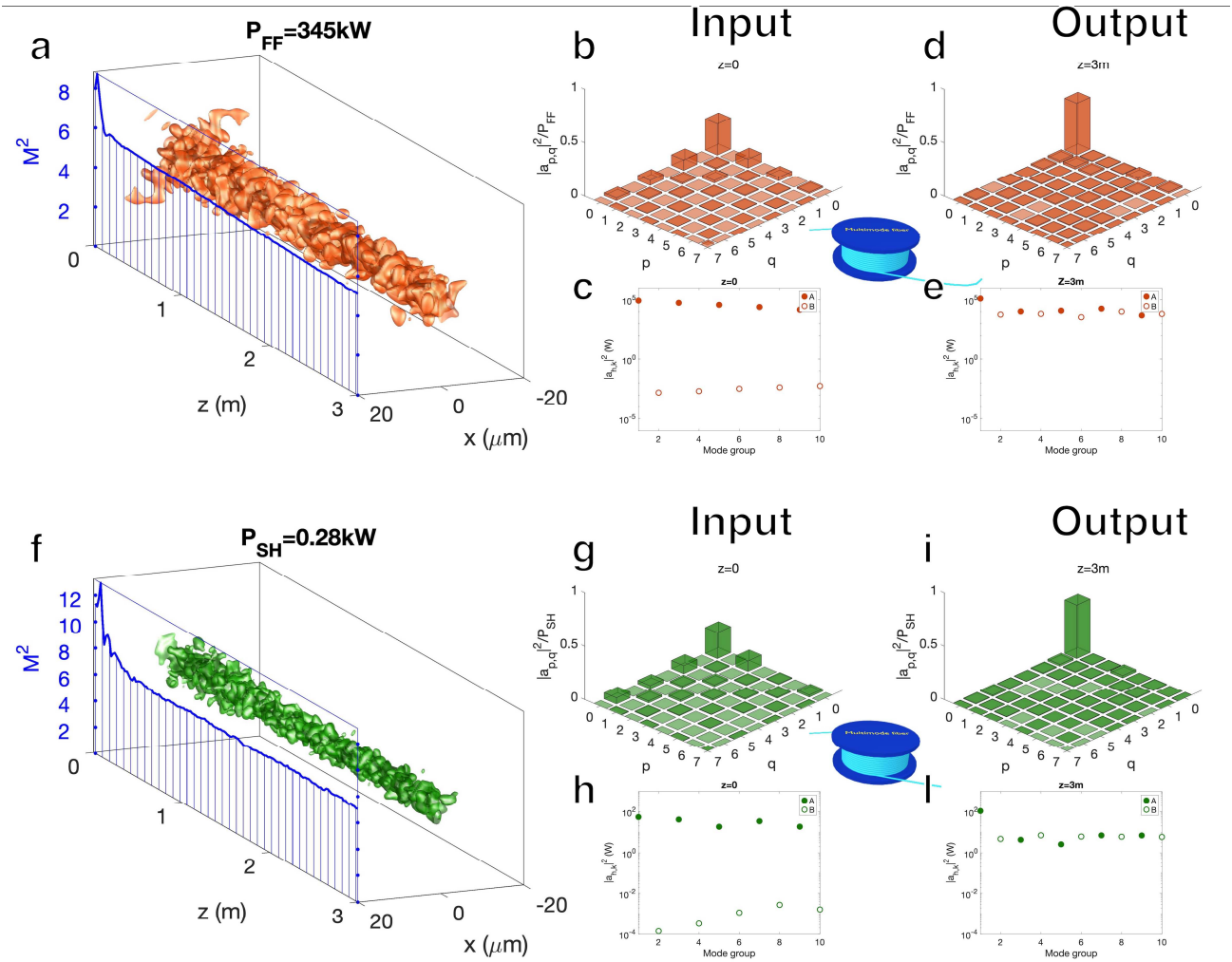

*Figure 4: Numerical simulations of BXC with temporal walk-off for FF and SH pulses; a, f: 3D evolution of the FF and SH beam with calculated M² coefficient; b, d and g, i: modal decomposition comparison between input and output of FF and SH beams, respectively; c, e and h, I: modal decomposition showing the difference in behaviour between odd(A) and even(B) modes for BXC with temporal walk-off, for two different propagation distances (0 m and 3 m).*

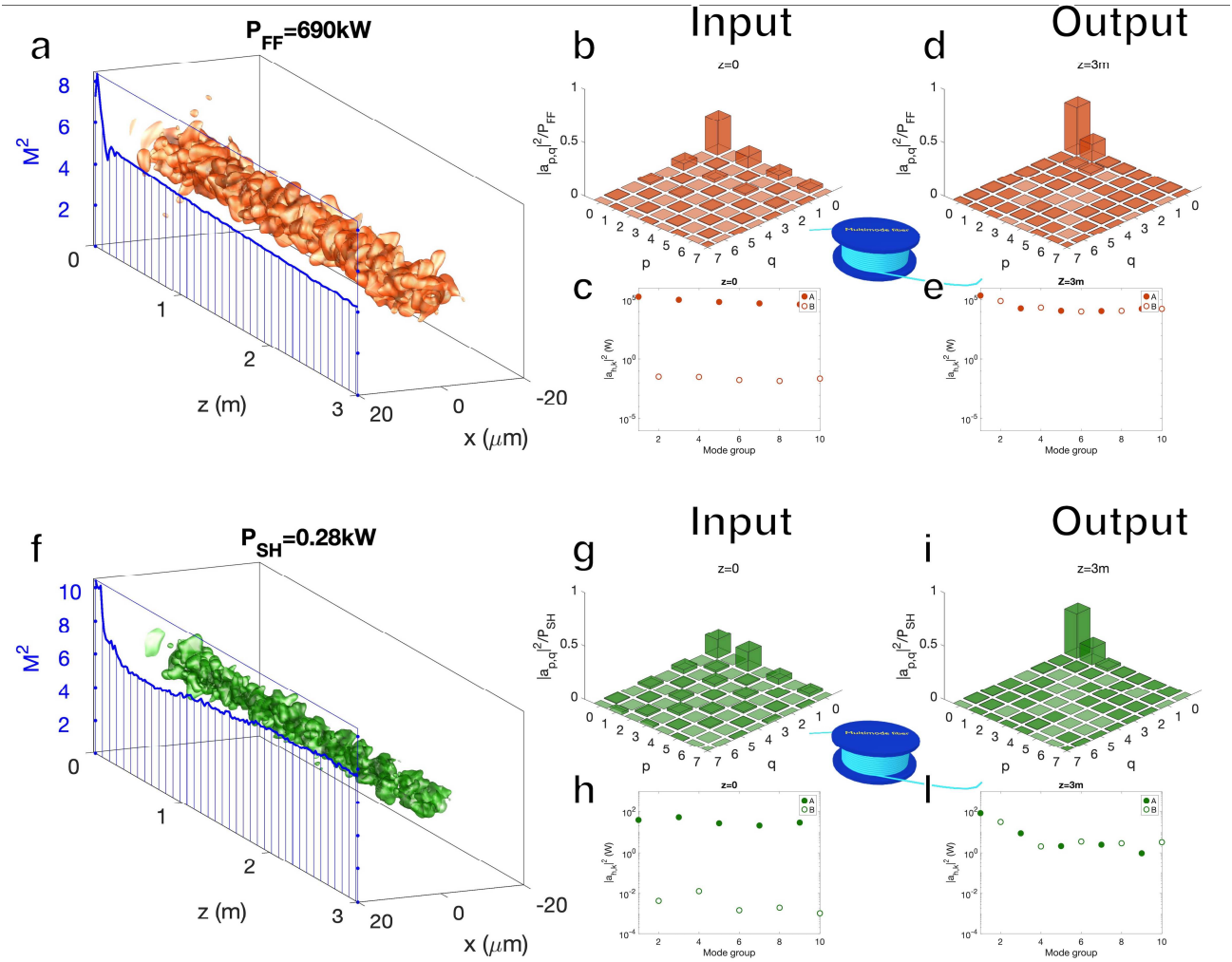

*Figure 5: Numerical simulations of BXS with temporal walk-off for FF and SH pulses; a, f: 3D evolution of the FF and SH beam with calculated M² coefficient; b, d and g, i: modal decomposition comparison between input and output of FF and SH beams, respectively; c, e and h, I: modal decomposition showing the difference in behaviour between odd(A) and even(B) modes for BXS with temporal walk-off, for two different propagation distances (0 m and 3 m).*

## DISCUSSION

In this work, we introduce and experimentally demonstrate the intriguing possibility of controlling the brightness of a weak beam via its interaction with an intense light beam at a different frequency, mediated by OIMC and XPM. This method adds an additional degree of freedom in the spatial BSC process that has been so far observed in multimode fibres. Cross-beam interaction is a purely conservative process, occurring without mutual amplification or absorption. It significantly expands the domain of application beam cleaning processed in Kerr media by enhancing the nonlinear interactions between copropagating waves. The modal overlap between wavelengths, which depends on the transverse position in the fibre core (see Figure 1), enables the control over the modal content of the weak SH beam. This is achieved via the spatiotemporal distribution of the pump intensity, which induces XPM-induced phase-matching for the spatial OIMC processes for the weak SH beam. The influence of XPM within a parametric FWM system is well established, and it was earlier demonstrated in microstructure fibres [44] (see Supplementary Information SI1).

Consequently, the power of the FF beam can control the rate and direction of energy transfer between modes in the weak SH. Under the combined self-imaging and Kerr effects, beam propagation in a GRIN optical fibre is longitudinally modulated by a transient index grating, which facilitates phase-matching of degenerate spatial IM-FWM processes, provided the peak power is sufficient. This phase-matching, induced by longitudinal intensity modulation, can also occur between waves at different frequencies, leading to geometric parametric instabilities [15], or between spatial modes, leading to the spatial BSC [16]. By coupling a high-power pump beam with a weak signal beam, one may control the efficiency of nondegenerate OIMC within each beam and change the direction of the energy flow among its modes, leading to either improvement or degradation of a laser beam quality. It should also be emphasized that the FF beam acts as an ultrafast diffuser on the SH beam, increasing or decreasing the energy transfer dynamics between modes depending on its initial peak power. This spatial energy redistribution in the SH beam can be associated with the transient evolution of its out-of-equilibrium entropy as a function of the FF peak power (see Supplementary Information SI4) [41]. Numerical simulations show that, the higher the slope of the entropy of the SH, the more effective the energy dispersion on the other transverse modes is, which increases the M² coefficient and decreases the beam quality. By pushing cross-cleaning to its limit, we numerically predict a complete transfer of energy from a weak speckle beam to a single transverse mode using a high-power FF speckle beam (see Supplementary Information SI5)

As a dramatic manifestation of the possible consequences of XPM-based control of different laser beams in MMFs, we experimentally demonstrated the ability to enhance or suppress the Raman cascade of a visible radiation, generated by SH of another infrared beam, based on the mere control of its input coupling condition into the fibre. We envisage that these findings will have many far-reaching impacts on applying nonlinear multimode fibres for high-power laser beam technologies.

**METHODS**

**Experiments.** The detailed experimental setup is presented in Figure 1(a). It consists of a femtosecond-picosecond laser source delivering pulses between 250 fs and 100 ps (1030 nm), and a laser source with a pulse duration of 650 ps. The initial beam is separated into two parts using a polarizer and a half-wave plate. The first beam is converted into a BBO crystal cut for SH generation, and the remaining part of the infrared pump is removed using a BG18 filter. The second part of the pump beam is attenuated by a passive filter and resynchronized with the SH of the first beam by means of two delay lines placed on each arm of the Mach-Zehnder interferometer. A wideband dichroic plate is used. The two beams are temporally and spatially superimposed at the fibre input with linear polarization states by using a half-wave plate at the FF wavelength placed before the coupling lens. The fibre excitation process is carried out using a converging lens with a focal length of 50 mm. The output beams are analysed by two CCD cameras, allowing near- and far-field recordings for both beams. The first fibre is a 10 m long GRIN 50/125 µm multimode guide. A second GRIN fibre, 50 m long with a 27 µm core diameter, is instead used for Raman cascade generation. The images on Figures 3(f), 3(g), 3(h) and 3(i) are obtained by spatio-spectral analysis, using an optical grating to disperse the light in the horizontal dimension and reveal the far-field image of the beam in the vertical plane (Figure 3(e)).

**Numerical simulations:** For the spatial propagation in both FF beam and SH, we solved the GNLSE by considering an input Gaussian beam with a FWHM of 35 $\mu m$ for both the SH and the FF. This simulation considers diffraction, material dispersion, waveguide contribution and Kerr effect ($n_2 = 3.2 \cdot 10^{-20} m^2/W$ and Raman fraction $f_r = 0.18$). The simulation considers both polarizations coupled incoherently by means of the Kerr effect along propagation. To achieve the BXC and BXS both beams are propagated simultaneously and the NLSE for the SH was modified to include the cross term as follows:

$$\begin{cases} \frac{\partial A_{SH}}{\partial z} = i \frac{\nabla_\perp^2 A_{SH}}{2k_{SH}} + i \frac{(k^2(x,y)-k_{SH}^2)}{2k_{SH}} A_{SH} + i\gamma_{SH}(|A_{SH}|^2 + 2|A_{FF}|^2)A_{SH}, \\ \frac{\partial A_{FF}}{\partial z} = i \frac{\nabla_\perp^2 A_{FF}}{2k_{FF}} + i \frac{(k^2(x,y)-k_{FF}^2)}{2k_{FF}} A_{FF} + i\gamma_{FF}(|A_{FF}|^2 + 2|A_{SH}|^2)A_{FF} \end{cases} \quad (2)$$

Where $A_{SH}$ and $A_{FF}$ denote the complex field evelopes at SH and FF respectively and $\gamma_i = \frac{(1-fr)n_2\omega_i}{c}$. The first term in the equation denotes the diffraction of the beam, the second one the waveguide condition and the third term accounts for the non-linear effects. To simulate the walk-off decoupling of the beams without including the temporal dimension in the simulation (to lighting the computational effort), after 3 m the cross terms between SH and FF were eliminated. It is also worth noting that in the FF equation, the cross term with the SH ($|A_{SH}|^2$), could be omitted as the $I_{FF} \gg I_{SH}$. While the intensity of the SH was kept constant at $I_{SH} = 2 GW/cm^2$ and the FF was increased until $I_{FF} = 1 TW/cm^2$. The guide specifications were those of a standard GRIN fibre 50/125 $\mu m$ with $n_{core} = 1.47$ and $n_{clad} = 1.457$, length of the fibre 6m. The numerical simulation was performed with a z-step of 12 $\mu m$, a spatial window of 90 $\mu m$ and resolution on x/y of 0.35 $\mu m$.

The modal decomposition performed both in the experimental and the simulations was done by means of a stochastic parallel gradient descent optimization algorithm [45]. The optimization was supervised by "epochs" of 100 times the number of modes considered, stopping criteria was either 100 "epochs" either an increase in the correlation experimental-simulation lower than $10^{-4}$. For the mode decomposition of the simulated data, the far field requirement was omitted as the phase of the beam was known.

**Acknowledgements**
VC thanks the National Research Agency under the Investments for the future programme with the reference *ANR-10-LABX-0074-01 Sigma-LIM, the ANR-23-CE08-0021-02, ANR-21-ESRE-0007 ADD4P, ANR-24-CE4+6-7295-01-ACT, and the EIC Pathfinder project "Multiscope" N°: 101185664. YA acknowledges receiving a postdoctoral fellowship (ED481D-2024-001) from the Xunta de Galicia (Spain). K. Krupa acknowledges Narodowe Centrum Nauki (OPUS 2023/49/B/ST7/01482) and Narodowa Agencja Wymiany Akademickiej (BPN/BIT/2024/00035). This work received funding from European Research Council (ERC) under the European Union's Horizon 2020 research and innovation programme under grant agreement No.950618 (STREAMLINE). B.W. also acknowledges the support of the Région Nouvelle Aquitaine (SPINAL Project).*

**Author contributions**
T. M., Y. A., V. C., K.K. and W. A. G. carried out the experiments. A. T., Y. A., M. F., S. W., A. P., and F. M. realized numerical simulation and proposed theoretical interpretation. All authors analysed the obtained results, and participated in the discussions and in the writing of the manuscript.

**Additional information**

**Competing financial interests**
The authors declare no competing financial interests.